\RequirePackage{xcolor}
\documentclass[journal,twoside,web]{ieeecolor}
\usepackage{generic}
\usepackage{cite}
\usepackage{amsmath,amssymb,amsfonts}
\usepackage{algorithmic}

\usepackage{graphicx}
\usepackage{textcomp}
\usepackage{tabularx}
\usepackage{booktabs}
\usepackage{verbatim}
\usepackage{multirow}
\usepackage{hyperref}
\usepackage[normalem]{ulem}

\newcommand{\revised}[1]{\textcolor{blue}{#1}}

\newcommand{\niu}[1]{\textcolor{blue}{#1}}
\definecolor{revcolor}{rgb}{0,0,1}
\definecolor{revcolor}{rgb}{0,0,0}
\renewcommand{\revised}[1]{#1}

\renewcommand{\niu}[1]{#1}
\renewcommand{\sout}[1]{}

\usepackage{etoolbox}
\makeatletter
\@ifundefined{color@begingroup}%
{\let\color@begingroup\relax
	\let\color@endgroup\relax}{}%
\def\fix@ieeecolor@hbox#1{%
	\hbox{\color@begingroup#1\color@endgroup}}
\patchcmd\@makecaption{\hbox}{\fix@ieeecolor@hbox}{}{\FAILED}
\patchcmd\@makecaption{\hbox}{\fix@ieeecolor@hbox}{}{\FAILED}

\def\BibTeX{{\rm B\kern-.05em{\sc i\kern-.025em b}\kern-.08em
    T\kern-.1667em\lower.7ex\hbox{E}\kern-.125emX}}
\begin{document}
\title{Explainable Diabetic Retinopathy Detection and Retinal Image Generation}
\author{Yuhao Niu, Lin Gu, Yitian Zhao, and Feng Lu, \IEEEmembership{Member, IEEE}
\thanks{This work was supported by National Natural Science Foundation of China (NSFC) under Grant 61972012 and JST, ACT-X Grant Number JPMJAX190D, Japan.}
\thanks{Y. Niu and F. Lu are with State Key Laboratory of VR Technology and Systems, School of Computer Science and Engineering, Beihang University, Beijing, China(e-mail: niuyuhao@buaa.edu.cn; lufeng@buaa.edu.cn). And F.Lu is also with Peng Cheng Laboratory, Shenzhen, China. }
\thanks{L. Gu is with RIKEN AIP, Tokyo, Japan, and also with University of Tokyo, Tokyo, Japan (e-mail: lin.gu@riken.jp).}
\thanks{Y. Zhao is with Cixi Instuitue of Biomedical Engineering, Ningbo Institute of Industrial Technology, CAS, Ningbo, China (e-mail: yitian.zhao@nimte.ac.cn).}
\thanks{Corresponding Author: Feng Lu (lufeng@buaa.edu.cn).}}

\maketitle

\begin{abstract}
Though deep learning has shown successful performance in classifying the label and severity stage of certain diseases, most of them give few explanations on how to make predictions. Inspired by Koch's Postulates, the foundation in evidence-based medicine (EBM) to identify the pathogen, we propose to exploit the interpretability of deep learning application in medical diagnosis. By determining and isolating the neuron activation patterns on which diabetic retinopathy (DR) detector relies to make decisions, we demonstrate the direct relation between the isolated neuron activation and lesions for a pathological explanation. To be specific, we first define novel pathological descriptors using activated neurons of the DR detector to encode both spatial and appearance information of lesions. Then, to visualize the symptom encoded in the descriptor, we propose \emph{Patho-GAN}, a new network to synthesize medically plausible retinal images. By manipulating these descriptors, we could even arbitrarily control the position, quantity, and categories of generated lesions. We also show that our synthesized images carry the symptoms \revised{\sout{that are}} directly related to diabetic retinopathy diagnosis. Our generated images are both qualitatively and quantitatively superior to the ones by previous methods. Besides, compared to existing methods that take hours to generate an image, our second level speed endows the potential to be an effective solution for data augmentation.

Code \revised{is available} at \url{https://github.com/zzdyyy/Patho-GAN}.
\end{abstract}


\section{Introduction}
	
\IEEEPARstart{D}{eep} learning has become a popular methodology in medical imaging analysis such as diabetic retinopathy (DR) detection~\cite{Gulshan2016jama} and skin cancer classification~\cite{esteva2017nature}. Though these algorithms have shown high accuracy in classifying specific disease labels or regressing severity stages, most of them lack the ability to explain the decision, a common problem that haunts deep learning community. Interpretability is especially imperative for medical image application, as physicians or doctors rely on medical evidence to determine whether to trust it or not. Otherwise, a simple adversarial attack ~\cite{PR20_AdversarialAttack} may result in irreversible damage. Similar limitation also echoes in other critical applications, like autonomous vehicle ~\cite{tian2018deeptest} and face recognition~\cite{WillifordMB20Explainable}, \textit{etc.}

%and there are more and more research on explainable deep learning.

In this paper, we propose a novel technique inspired by  Koch's Postulates to give some insights into how convolutional neural network (CNN) based medical imaging detector makes decisions. In particular, we take the diabetic retinopathy (DR), a common cause of vision loss among people with diabetes, for example. Note that not limited to DR detector~\cite{oO2016detector}, this interpretation strategy could also be extended to other deep learning based medical imaging models.

\begin{figure*}[h!]
	\begin{center}
		\includegraphics[width=0.9\linewidth]{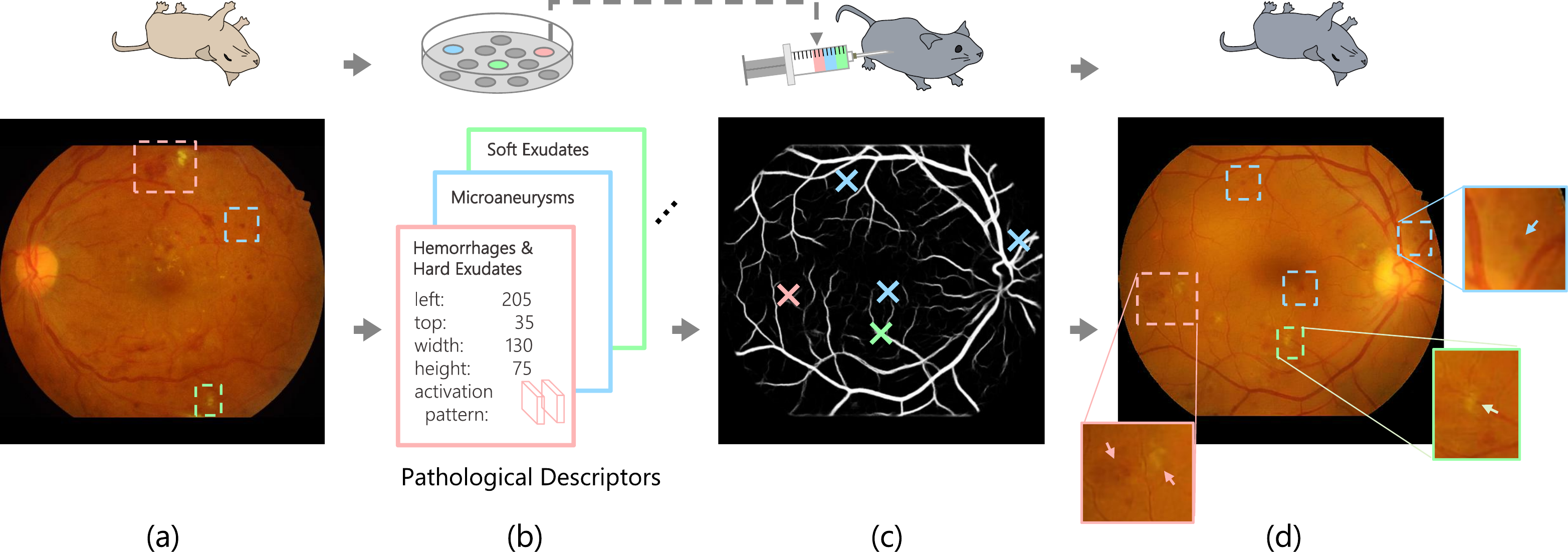}
	\end{center}
	\caption{Koch's Postulates is a fundamental criteria to determine the association between pathogen and the disease. As the top row shown, it states that (a) The pathogen must be found in diseased subjects but not in healthy ones; (b) The pathogen must be isolated and grown in pure culture; (c)\&(d) After being injected into healthy subject, the cultured pathogen should result in the symptom of diseases after, and the pathogen isolated again is the same as the injected one. The methodology of this paper is an analogy to Koch's Postulates. (a) The input reference image with several lesions such as hemorrhages and hard exudates in red box, microaneurysms in blue box, and soft exudates in green box. (b) Extract pathological descriptors that encode the lesion neuron pattern from an image like separating pathogen. (c) Apply descriptors at the designated positions on a binary vessel segmentation like injecting purified pathogen into healthy subject. (d) The synthesized image demonstrates the expected types of symptoms in the designated positions and quantity.}
	\label{fig:koch}
\end{figure*}

We at first define and extract pathological descriptors (Fig~\ref{fig:koch}.(b)) that encode the neuron activation directly related to the prediction of the DR detector~\cite{oO2016detector}. Isolating the directly relevant neurons, usually thousands, from  millions of neurons in neural network is like separating the potential pathogen from diseased organism. Koch's Postulates, the foundation of evidence-based medicine (EBM), could indirectly determine the property of the pathogen by observing the caused symptom. Specifically, as illustrated in the top row of Fig~\ref{fig:koch}, the candidate pathogen is at first purified from a variety of microorganisms. If the healthy subjects injected with this pathogen consistently show the symptom, Koch's  Postulates could claim the association between the pathogen and disease. Similarly, we \emph{inject} pathological descriptors into the binary vessel segmentation (Fig~\ref{fig:koch}.(c)) to synthesize the medical plausible retinal image with the expected lesions. We achieve this by proposing \emph{Patho-GAN}, a novel generative adversarial network (GAN) \revised{\sout{,  as illustrated in Fig~\ref{fig:pipeline}}}. With given pathological descriptors and binary vessel segmentation, the generated image (Fig~\ref{fig:koch}.(d))  exhibits the expected symptoms with specified quantity at the designated positions, such as hemorrhages and hard exudates in the red box, microaneurysms in the blue box, and soft exudates in the green box. 

%apply DR detector~\cite{oO2016detector} on the  reference image (Fig~\ref{fig:koch}.(a)) and extract the pathological descriptor (Fig~\ref{fig:koch}.(b)) that encodes the neuron activation directly related to prediction. Picking thousand out of  millions of neurons in neural network is like separating the potential pathogen. Koch's Postulates claims that the property of pathogen, though invisible for naked eye, could be determined by observing the arose symptom after injecting it into subject.

 Experiments on various datasets demonstrate that our generated images are both qualitatively and quantitatively superior to existing methods~\cite{zhao2018synthesizing}~\cite{Niu19AAAI}. Since our descriptors are lesion-based with spatial coordinates, we can arbitrarily manipulate these descriptors in position and quantity to generate the images with controlled type and quantity of symptoms. Fed with these generated images, the DR detecting algorithm~\cite{oO2016detector} also reports the  prediction consistent to the diagnose criteria. This shows our synthesized images carry the symptoms \revised{\sout{that are}} directly related to diabetic retinopathy diagnose. Moreover, the great diversity and high generating speed make \emph{Patho-GAN} a potential data augmentation method for medical image analysis.

In conclusion, our main contributions are mainly three-fold:

\begin{enumerate}
\item We define novel pathological descriptors that encode the key neuron activation \revised{\sout{that are}} directly related to the disease prediction. Each descriptor could be associated with individual specific lesion. 
\item Inspired by Koch's Postulates, we propose a novel interpretability strategy to visualize the pathological descriptors by synthesizing fully controllable pathological images. The synthesized images are qualitatively and quantitatively superior to previous methods.
\item With our pathological descriptors and \emph{Patho-GAN}, we could generate medical plausible pathology retinal images at a greater variety, quality, and speed, which enable \emph{Patho-GAN} to help improve the performance of medical tasks.
\end{enumerate}

This article is an extension work of our previous conference publication~\cite{Niu19AAAI}, here we have obtained the following improvements:
\begin{enumerate}
	\item\textbf{Novel Encoding Strategy.} We have improved our method on lesion locating and descriptor encoding. This enhanced its ability to represent a greater variety of lesions.
	\item \textbf{Photo-realism Enhancement.} We have improved the synthesis quality by training our Patho-GAN on the dataset of DR image, rather than healthy ones. Meanwhile, our method could get rid of artifacts like the checkerboard effect in ~\cite{Niu19AAAI}.
	\item \textbf{Speed Boost.} Our previous method used the descriptors in a restrictive loss function. For each new image, the whole model needed to be retrained. Here, by formulating the descriptors as input and generate the image in a forward pass, we can generate images much faster than ~\cite{Niu19AAAI}. This makes \emph{Patho-GAN} a potential data augmentation method for medical image analysis.
\end{enumerate}

\section{Related Works}
\subsection{Diabetic Retinopathy Detection}
\label{sec:DR detector}

Diabetic retinopathy (DR) is a common disease cause of vision loss or even blindness among people with diabetes, which affects 347 million people~\cite{danaei2011national}. Human ophthalmologists rely on the type and amount of related lesions, let us say, counting the number of microaneurysms, to  grade the DR severity. As a result, automatic detection is particularly needed to reduce the workload of ophthalmologists, and slow down the progress of DR by performing early diagnose on diabetic patients~\cite{Gulshan2016jama}.

%Then, another deep learning method~\cite{Gulshan2016jama}, trained on  128,175 images, achieved a high sensitivity and specificity for detecting diabetic retinopathy.

 In 2015, a Kaggle competition~\cite{kaggle2015diabetic} was organized to automatically classify retinal images into five stages according to \textit{International Clinical Diabetic Retinopathy Disease Severity Scale}~\cite{aao2002drscale}. Not surprisingly, all of the top-ranking methods were based on deep learning. However, though achieving high sensitivity and specificity, these deep learning based methods lack the  intuitive explanation for the decision. The recent methods~\cite{Yang17MICCAI,wang2017zoom,Lin18AFN} shifted the focus to locate the lesion position with a weakly supervised learning framework. However, these methods often relied on a large training set of lesion annotations from professional experts.
 
To explore the intuitive explanation, we propose a novel  pathological descriptor encoding DR detector's activated neurons directly related to the pathology. For the sake of generality, we select o\_O DR detector, a CNN based method within the top-3 entries on Kaggle's challenge. Even by now, the performance of o\_O is still competent to the latest method~\cite{Lin18AFN}. %This method is trained and tested on the image level DR severity labeling. 

\subsection{Explainable Deep Learning}
 
\revised{
%In sensitive and critical applications such as medicine, banking and self-driving automobiles, explanations are needed to enhance decision reliability, algorithm fairness, and ensure network performance. 
Explainable deep learning has now emerged as an important research to open the black-box of neuron networks. Existing researches mainly focus on answering the following three questions:~\cite{gilpin2018explaining,DBLP:journals/entropy/LinardatosPK21,Li20SCOUTER}.
}

\revised{\textbf{How is the input processed by the network?} Many efforts are devoted to training simple proxy models and emulating the network's \textbf{processing} of data to discover connections between output and input. For example 
LIME~\cite{ribeiro2016should} approximates a classifier locally with interpretable linear model. The proxy model is able to identify influential input regions of a network across various model and applications, and thus faithfully explains the original classifier. 
}
%\lin{Please add at least one paragraph on this part, citing at least 5 papers}
\niu{CRED~\cite{sato2001rule} and recently extended method DeepRED~\cite{zilke2016deepred} choose decision trees as proxy model to imitate the inference process of neural networks. More recent works~\cite{hinton2017distilling,tan2018distill,wu2018beyond} try to distill knowledge from deep models into trees, but the proxy model are more used for classification and less human-interpretable. Zhang \textit{et al.}~\cite{zhang2019interpreting} propose to build semantic decision tree that encodes all potential decision modes in a coarse-to-fine manner.}

\revised{\textbf{How to design explainable systems?} There are also researches designing self-explainable structures to achieve \textbf{intrinsic} interpretability. 
The structures include attention based CNN layers, modules, or other particularly designed pipelines. }
\niu{Li \textit{et al.}~\cite{Li20SCOUTER} propose a slot attention-based classifier for transparent yet accurate classification, which provide positive or negative explanation for a certain category. Chen\textit{et al.}~\cite{Chen19TLLT} insert a layer of prototype correlation, making decisions by finding prototypical parts and combining evidence like human experts. 
}
%For example, they use attention based CNN layers or modules~\cite{wang2017zoom,Li20SCOUTER}, or other particularly designed pipelines~\cite{Lin18AFN, Chen19TLLT}. %Unfortunately, there is often a trade-off between intrinsic interpretability and model performance like accuracy. 
\revised{Zhang \textit{et al.}~\cite{zhang2020Interpretable} disentangle ``the mixture of pattern'' in same filter by a specific loss restriction and build an interpretable object classification network.
}%
%\lin{Similarly, extend this to a paragraph, at least citing 5 more papers}
\niu{These self-explainable technique has been applied to DR detection. Zoom-in-Net~\cite{wang2017zoom} uses generated attention map to automatically discover suspicious regions for a double check, some of which are meaningful lesions. AFN~\cite{Lin18AFN} presents Center-Sample detector in the pipeline to evidently find lesions in retinal images through feature space clustering, and fuses lesion map with original retinal image in Attention Fusion Network. They achieve state-of-the-art accuracy while providing enough explanations.}

\revised{\textbf{What information does the network contain?} Given an existing successful neuron network, how to decode its layer- or neuron-level \textbf{representation} has attracted much attention.~\cite{gilpin2018explaining,Li20SCOUTER}.}
%in a pre-trained network, usually by various visualization methods~\cite{gilpin2018explaining,Li20SCOUTER}.
%including perturbation, gradient-based back-propagation, optimization.
%The explanations are either given by the perturbation experiments, or gradient-based visualizations like Grad-CAM~\cite{Selvaraju20GradCAM} and guided back-propagation~\cite{zeiler2014visualizing}.
\niu{\emph{Perturbation-based} methods perturb the input data to analyze the consequent change in the neuronal activity. Occlusion~\cite{zeiler2014visualizing} occludes part of the input image with a sliding window, while RISE~\cite{DBLP:conf/bmvc/PetsiukDS18} applies random masks on the input. By modifying input and observing how the neurons change, these methods link neuronal activation with specific input patterns. \emph{Optimization} methods~\cite{mahendran2016visualizing,nguyen2016multifaceted} directly optimize input image to maximize specific neurons. The optimized image shows what pattern can most activate the neurons.}

\niu{
\emph{Back-propagation} methods like CAM~\cite{zhou2016learning}, GradGAM~\cite{Selvaraju20GradCAM}, GradCAM++~\cite{chattopadhay2018grad} and DeepLIFT~\cite{shrikumar2017learning} produce heat maps with (gradient-based) back-propagation to visualize the discriminant regions for class prediction.}
\revised{
A back-propagation method~\cite{zeiler2014visualizing} uses Deconvolution to restore images from features in different layers, giving insight on layer-wise interpretation in deep networks. A variant method Guided BackPropagation~\cite{springenberg2014striving}, also known as guided saliency, can be used for visualizing features learned by CNNs, which can also be applied to a broad range of network structures.
}

\revised{Besides the attributive explanation for one example, there are more methods solving new multi-example tasks\cite{goyal2019counterfactual,wang2020scout}. 
}%\lin{Please use at least two paragraphs to describe this part}

\revised{To enhance the mutual understanding between the community of artificial intelligence and clinical medicine, we propose a novel framework to explain the neuron-level representation following the methodology of  evidence-based medicine (EBM). Though the framework is evaluated on Diabetic Retinopathy Detection~\cite{kaggle2015diabetic}, it could be easily applied to explain other medical deep learning algorithms. We at first automatically identify the neuron representation that encodes the symptom information that contributes to the disease classification. Then by synthesizing lesions in encoded DR descriptors, we are able to associate pathological lesions with related neuron activation patterns.}

%\revised{Our proposed method explains the neuron-level representation in DR detection network through GAN-based visualization. We first locate lesions through deconvolution back-propagation~\cite{zeiler2014visualizing}. And then by synthesizing lesions in encoded DR descriptors, we associate pathological lesions with related neuron activation patterns.  }

\subsection{Generative Adversarial Networks}

Generative Adversarial Networks (GANs)~\cite{goodfellow2014generative} were first proposed in 2014, adopting the idea of zero-sum game, to generate realistic images. Subsequently, CGANs~\cite{mirza2014conditional} attempted to use additional information to make the GAN controllable. More recent methods use GAN in image-to-image translation. Pix2pix~\cite{isola2017image} used the U-Net~\cite{ronneberger2015u} combined with adversarial training and achieved amazing results. CycleGAN~\cite{zhu2017unpaired} used two sets of GANs and added cycle loss to achieve style transfer on unpaired data.  These methods have been applied in many medical image processing tasks such as PET-CT translation, correction of MR motion artefacts and PET image denoising \cite{armanious2020medgan}. Besides, attempts of cross-modal translation~\cite{liang2020cpgan} will shed light on mutual understanding between medical text and images. %DCGAN~\cite{radford2016unsupervised} combined CNN with traditional GAN to achieve a shocking effect. %As shown in~\cite{yi2019generative}, GANs has been helped medical imaging in synthesis, reconstruction, segmentation, classification, etc.

%\textcolor{red}{This section needs more details, and should focus on GAN-based medical image analysis. In addition, I suggest merging II.C and II.B.}

\begin{figure*}[t]
	\begin{center}
		\includegraphics[width=\textwidth]{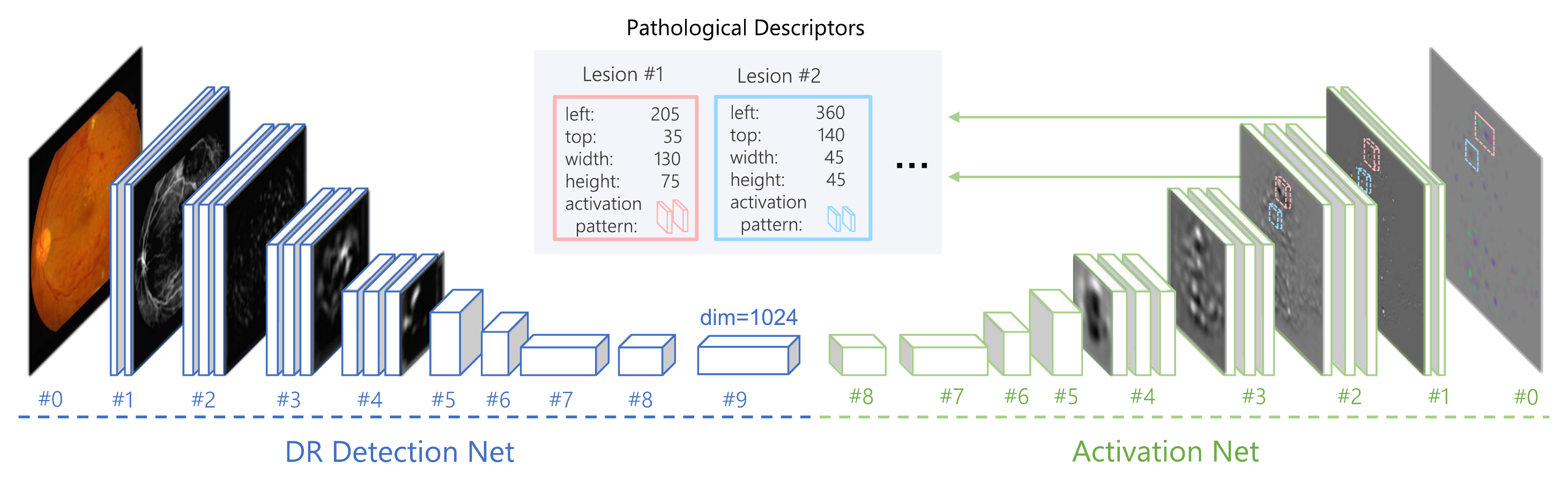}
	\end{center}
	\caption{The process of extracting pathological descriptors. First, a pathological reference image is fed into the off-the-shelf  \emph{detection net}. The extracted features are then back-propagate to the input pixel space through the symmetric \emph{activation net} to get activation projections, which indicate the locations and appearance of most lesions. Finally, the activation projections are cropped into small patches around the found lesions, forming pathological descriptors. Note that the skip connections from pooling layers (DR Detection Net) to (Activation Net) unpooling layers are not shown in this figure. }
	\label{fig:descriptor}
\end{figure*}
   	
\subsection{Synthesizing Biomedical Images}

%Probably most well-known efforts are the work~\cite{fiorini2014automatic}, the work~\cite{bonaldi2016automatic}, GENESIS~\cite{bower2015genesis}, NEURON~\cite{carnevale2006neuron}, L-Neuron~\cite{ascoli2000neuron} \textit{etc.}

Traditionally, the  biomedical images were synthesized through the medical and biological prior knowledge accumulated by humans. Combined with complex simulation methods, realistic results could be produced~\cite{fiorini2014automatic,bonaldi2016automatic,bonaldi2016automatic,bower2015genesis,carnevale2006neuron,carnevale2006neuron,ascoli2000neuron}. 	

With the development of generative models such as GAN and deep neural style transfer, Tub-sGAN~\cite{zhao2018synthesizing} began to synthesize realistic retinal and neuronal images in a data-driven way.
\revised{It synthesized images with style from a reference retinal image and with vessels from a binary segmentation map \sout{either manually annotated or automatically segmented}~\cite{Liu19MICCAI,guo2020saunet}}.
Although their generated images could show pleasant visual appearance, the diabetic retinopathy symptoms and retina physiological details are either lost or incorrect as verified by the ophthalmologists~\cite{Niu19AAAI}. \revised{And for a new reference retinal image,  it spends dozens of minutes in training new generator.}

%\revised{
%A more recent work DR-GAN~\cite{Zhou19DRGAN} can synthesize high-resolution DR images with pixel-wise vessel and lesion masks, while keeping a reasonable DR severity grade. Besides, there are experiments on how } However, this work relies on the lesion segmentation. Besides, it could not perform  the lesion-level explainable manipulation. In this paper, we propose a pathologically controllable method that can generate realistic retinal image with medical plausible symptoms.

%We  refer to DR-GAN for its training  dataset for photo-realism enhancement and its input scheme for speed boost. We refer to DR-GAN for its training dataset to get photo-realism enhancement and its input scheme to get speed boost.

\revised{
A more recent work, DR-GAN~\cite{Zhou19DRGAN}, can synthesize high-resolution DR images by introducing multi-scale spatial and channel attention module, while keeping a reasonable DR severity grade by specially designing the latent space and classification loss. \niu{DR-GAN is fed with retinal structure and lesion masks, as well as a latent code of given severity. It generates retinal fundus with lesions specified with mask and severity. } Different from DR-GAN that focus on improving the synthesizing quality through GAN, our proposed method is an interpretability work that explains ``how an image is being classified''. To achieve this end, we synthesize the descriptors automatically extracted from the DR detector and generate vivid lesions that are clinically related to DR.
Different from DR-GAN that relies on the input of lesion mask, our method automatically encodes the descriptors from DR images without explicitly knowing these leision information. Our Patho-Gan could then synthesise these lesions in new images. With the descriptors encoding the neuron representation, we could arbitrarily manipulate the lesion types, number, and locations when generating the image.
}

\section{Pathological Descriptor}

In this section, we will describe how to extract lesion based pathological descriptor to encode the activated neurons of Diabetic Retinopathy (DR) detector~\cite{oO2016detector}.

\subsection{DR Detection Network}
Here, we  briefly introduce o\_O DR detector~\cite{oO2016detector} used in this paper. It takes retinal fundus image $x$ as input  and outputs the 5 grades (0-4) diabetic retinopathy severity $s$.
As shown in the left part of Fig~\ref{fig:descriptor}, the DR detection network is stacked with several blocks. Each block consists of 2-3 convolutional layers and a pooling layer. As the number of layers increases, the network merges into a  $ 1\times 1 \times 1024$ \emph{bottleneck feature} $ F_9(x) $. 
Finally, $ F_9(x) $ is fed into a dense layer (not shown in the figure) to predict the severity label $ s=f(F_9(x))$. The DR detector is denoted as a function $s=\text{DR}(x)$.
% To add nonlinearity to the net and to avoid neuronal death, a leaky ReLU~\cite{maas2013rectifier} with negative slope 0.01 is applied following each convolutional and dense layer.

 The network is trained on Kaggle DR dataset~\cite{kaggle2015diabetic} with Nesterov momentum over 250 epochs. Data augmentation methods, such as dynamic data re-sampling, random stretching, rotation, flipping, and color augmentation, are all applied.

After the training, the DR detector have learned \textit{whether to diagnose an image as DR}. We will keep going and explore \textit{why the DR detector give such diagnosis}.

\subsection{Activation Network}

Among millions of neurons in the network, only thousands of them actually contribute to the \emph{bottleneck feature}'s activation and the final prediction. To explore the attributed activity of these neurons, we perform a back-propagation-liked procedure~\cite{zeiler2014visualizing} from the $1024$-dimensional \emph{bottleneck feature} to get \emph{activation projections} $ \{A_l(x)\}  $ for each DR detector's feature layer $ l $, which encodes the impact of each neuron to the bottleneck feature~\cite{Niu19AAAI}.  Thus, we back propagated through each layer to build an symmetric network which we called \emph{Activation Net}.

%build an  an symmetric network\emph{Activation Net}

%This procedure back propagated through each layer is an symmetric network which we called \emph{Activation Net}.

As shown in the right part of Fig~\ref{fig:descriptor}, our activation net is a reversed version of the DR detector with the replicated weights. For each layer in the detector, there is a corresponding reverse layer in the activation net, with the same configuration of strides and kernel size: (1) For a convolutional layer, the corresponding layer performs transposed convolution, which shares the same weights, except that the kernel is flipped vertically and horizontally; (2) For each max pooling layer, there is an unpooling layer that conducts a partially inverse operation, where the max elements are located through a skip connection (not shown in the figure) and non-maximum elements are filled with zeros; (3) For a ReLU function, there is also a ReLU in the activation net, which drops out the negative \emph{activation projection}; (4) The fully connection dense layer can be treated as a $1\times 1$ convolution. In the implementation, we use auto-differentiation provided in Tensorflow~\cite{abadi2016tensorflow} to conduct back-propagate for each layer. 

\subsection{Explanation of DR Detection}

\begin{figure}[t]
	\begin{center}
		\includegraphics[width=\columnwidth]{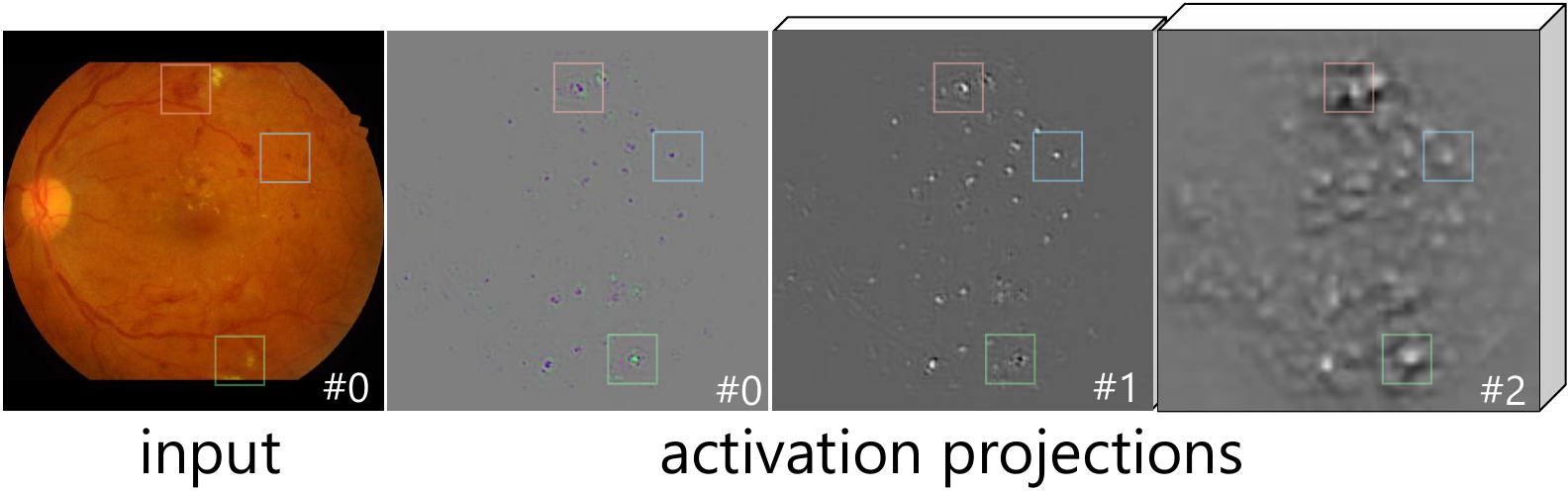}
	\end{center}
	\caption{We feed fundus image $x$ into the pipeline in Fig~\ref{fig:descriptor} and show their related activation projections $ A_l $ in layer $l$. }
	\label{fig:activation_projection}
\end{figure}

\begin{figure}[t]
	\begin{center}
		\includegraphics[width=\columnwidth]{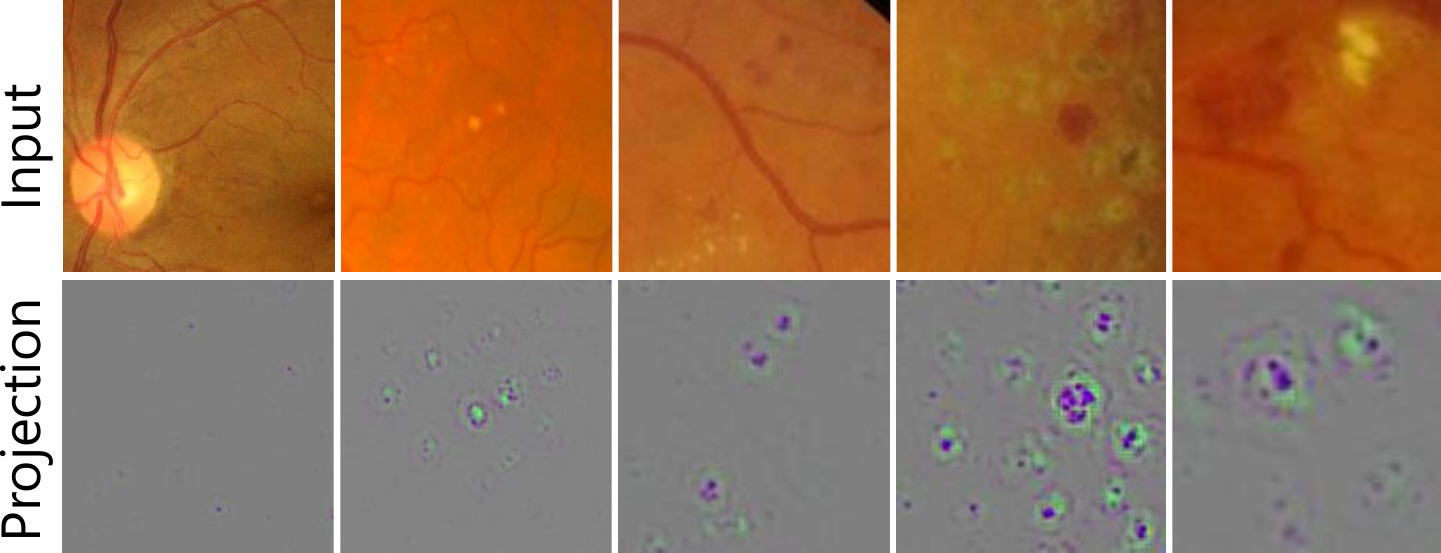}
	\end{center}
	\caption{Zoomed in fundus image  $ x $  and their activation projections $ A_0 (x) $ in layer \#0 . }
	\label{fig:detected_lesion}
\vspace{0.7em}
	\begin{center}
		\includegraphics[width=\columnwidth]{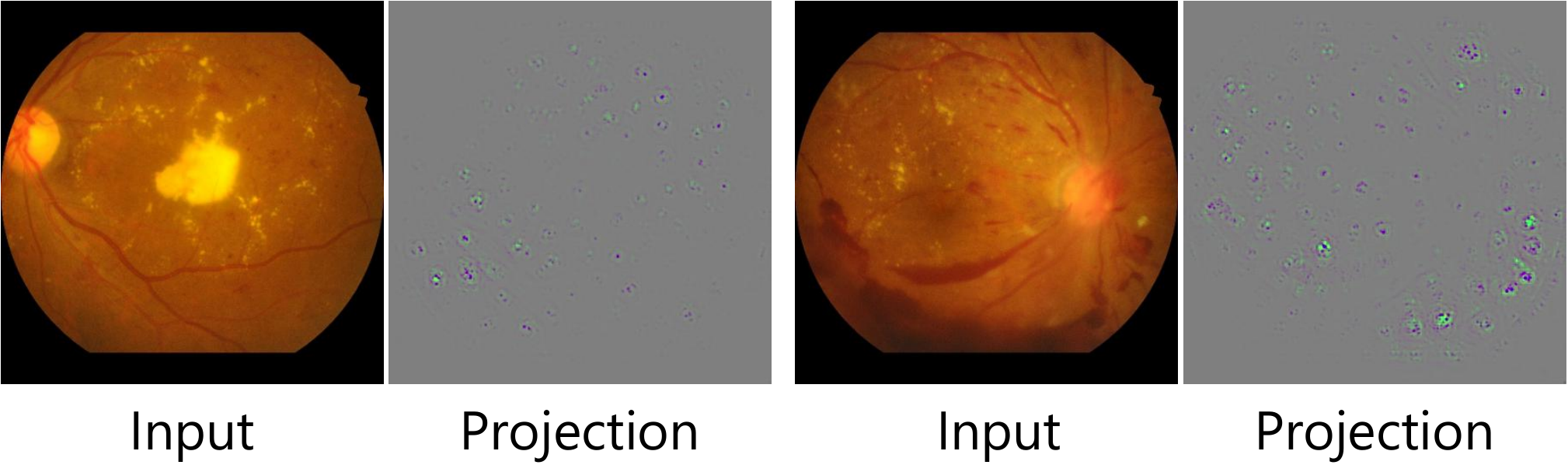}
	\end{center}
	\caption{For the very large-region exudates or hemorrhages, their activation projections are not obvious.}
	\label{fig:activation_projection_failure}
\end{figure}

Here in Fig~\ref{fig:activation_projection}, we illustrate an input image $x$ and its activation projections $\{A_l(x)\}$ in 3 different layers, \#0, \#1, and \#2. From the input layer \#0 to the activation projection layer \#0, only the lesion regions arise obvious activation (such as the hemorrhages in blue and pink boxes, and the exudates in green boxes). This locality allows us to associate each activated immediate surrounding in the $A_l(x)$ with  a specific DR lesion in $ x $.

%However, as shown in Fig~\ref{fig:activation_projection_failure}, the $ A_0 $ may fail to response to the large-region exudates or hemorrhages.

Fig~\ref{fig:detected_lesion} gives more examples on  various types of lesions along with their activation projections. We could observe that $ A_0 $ ignores the the normal physiological region and only responses on the abnormal regions with microaneurysms, hard exudates, hemorrhages and soft exudates. \niu{Very interesting, as as shown in Fig~\ref{fig:activation_projection_failure}, $ A_0 $ does not response to the large-region exudates or hemorrhages, because the existence of microaneurysms are sufficient for o\_O detector to make diagnose. This shows that o\_O detector can predict severity without  detecting large ones.}

%, only recognizing small ones.
%\niu{In fact, Fig.~\ref{fig:activation_projection_failure} actually shows that o\_O detector can predict severity without the ability to detecting large ones.}

The selective response of  activation projections on the lesion indicates that our method could spatially recognize meaningful lesion patterns without the pixel-wise  pathological annotation for training. In the meantime, the symptoms selected by the DR detector is also consistent with the ophthalmologists used for diagnose~\cite{Niu19AAAI}. Based on this observation, we aim to encode the neuron activation pattern that is directly related to the disease diagnose as an analogy of isolating the pathogen in Koch's postulates. 

%even trained with image-level severity labels and lacks pathological annotation,
%it can automatically learn to recognize meaningful lesion patterns, and use it in further diagnosis.
%Most of the time, our DR detector makes decisions depending on the same symptoms with ophthalmologists. But in extreme conditions like Fig~\ref{fig:activation_projection_failure}, it only sees small lesions, far from the actual performance of ophthalmologists.

\subsection{Retinal Pathological Descriptor}

As the neuron activation projection $ A_0(x) $ is spatially correlated with the retinal lesions in $ x $, we could define \emph{pathological descriptor} set $ \mathcal{D} = \{d_r\} $ to encode the coordinates, dimensions and activation patterns of individual lesion $ r $. These lesions  serve as the evidence  for both ophthalmologists and  DR detector to  make diagnosis.  

%When a retinal fundus $ x $ is fed into the pipeline in Fig~\ref{fig:descriptor}, the features and activation projections in layer $ l $ are denoted as $ F_l $ and $ A_l $, respectively.

\begin{figure}[t]
	\begin{center}
		\includegraphics[width=\columnwidth]{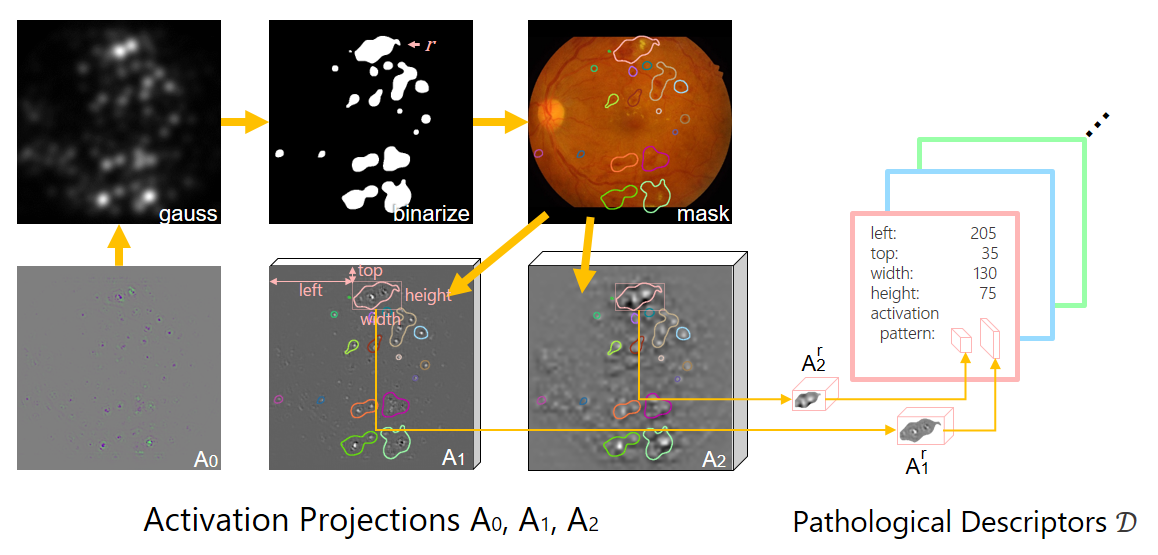}
	\end{center}
	\caption{Extraction procedure of pathological descriptors. }
	\label{fig:descriptor_extraction}
\vspace{0.7em}
	\begin{center}
		\includegraphics[width=\columnwidth]{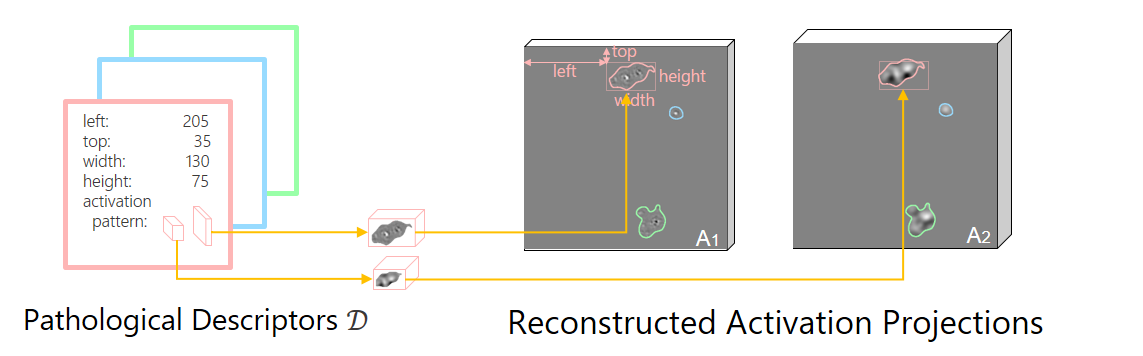}
	\end{center}
	\caption{The reconstruction of activation projection from pathological descriptors. }
	\label{fig:descriptor_reconstruction}
\end{figure}

Fig~\ref{fig:descriptor_extraction} illustrates how the pathological descriptors are extracted:
(1) To locate a lesion, we process the last layer activation $ A_0(x) $  by Gaussian blur with $ \sigma=10 $ before thresholding it with Otsu's method~\cite{otsu1979threshold}. Thus we get a binary mask where different lesions are separated. %For example, we denote the uppermost region as $r$.
(2) For each lesion, we enclose it with a minimal bounding box, like the pink one $r$ in Fig~\ref{fig:descriptor_extraction}. %The coordinate of top and left as well as the dimensions of width and height are then recorded.
(3) We extract vectors of activation projection $ A_l $ in the bounding box to get the activation pattern $ A_l^r $. As shown in  Fig~\ref{fig:descriptor_extraction}, we collect the neuron outputs $ A_1^r $ and $ A_2^r $ from activation projection layers $ A_1 $ and $ A_2 $ in the array of bounding box. %For example, here in the figure we are interested in 2 activation projection layers $ A_1 $ and $ A_2 $, we crop these two array at the pink lesion region $ r $ to get activation pattern $ A_1^r $ and $ A_2^r $. 
(4) With the bounding box coordinates, dimensions and activation patterns, we could define a pathological descriptor $ d_r=\left<\text{left}(r), \text{top}(r), \text{width}(r), \text{height}(r), A_1^r, A_2^r \right> $ for each lesion $r$. 
(5) Finally, by repeating step 2-4, we could get descriptors for every lesion in $ x $, forming a descriptor set $ \mathcal{D}(x) = \{d_r\} $.

In summary, for an fundus image $ x $, we can extract a set of descriptors $ \mathcal{D}(x) $, which encodes the coordinates, dimensions and activation patterns of lesions in $x$. Furthermore, descriptors $ \mathcal{D} $ can be used to reconstruct activation projections by inserting the encoded lesion associated neuron action into the activation maps (Fig.~\ref{fig:descriptor_reconstruction}).

\begin{figure}[t]
	\begin{center}
		\includegraphics[width=\columnwidth]{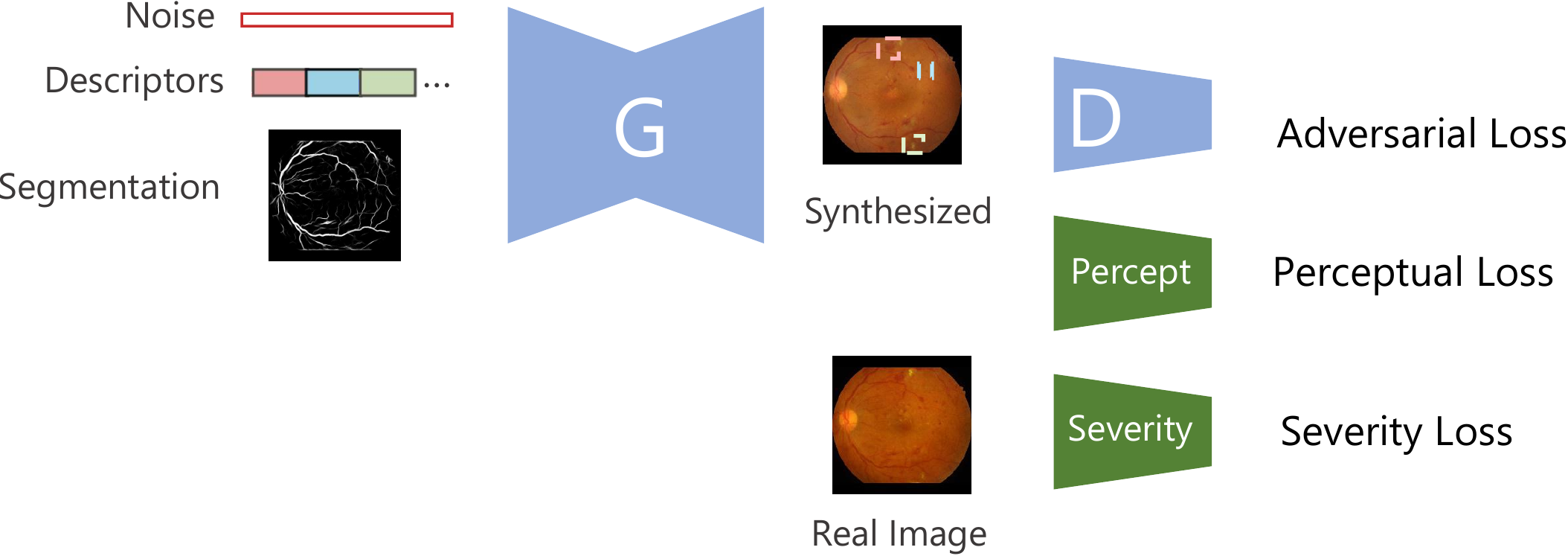}
	\end{center}
	\caption{\revised{Overview of Patho-GAN training. The Generator is restricted with 3 sub-networks and losses. Blue networks are trainable, while green network are pre-trained and fixed. }}
	\label{fig:pipeline_overview}
\end{figure}

\begin{figure*}[t]
	\begin{center}
		\includegraphics[width=\textwidth]{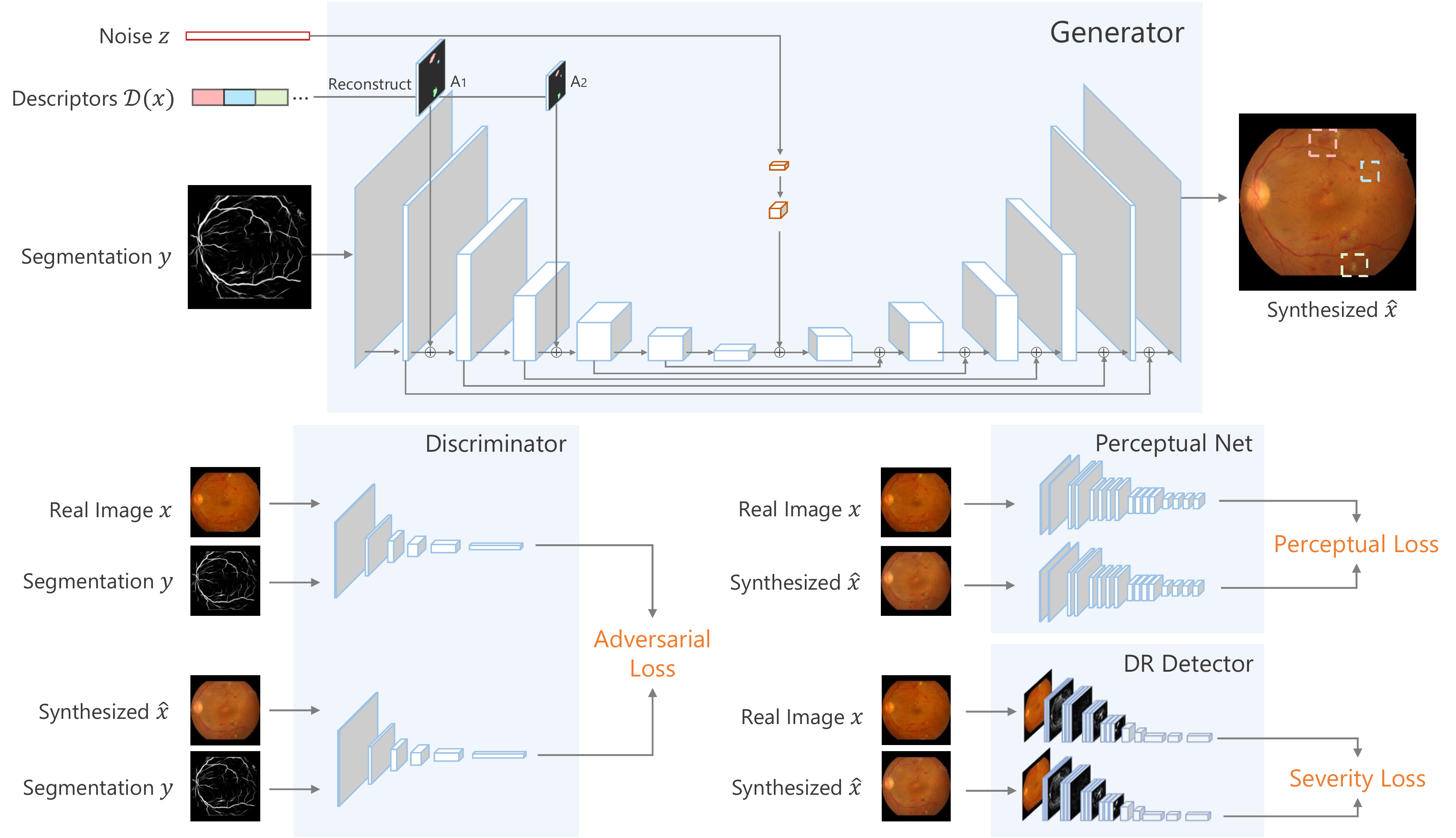}
	\end{center}
	\caption{The architecture and data flow of Patho-GAN ($ \oplus $ means concatenation). We design three types of loss: Adversarial Loss, Perceptual Loss, and Severity Loss in training phase. After training, the generator could directly synthesize retinal fundus with lesions given the input of pathological descriptors and binary vessel mask.}
	\label{fig:pipeline}
\end{figure*}

\newpage

\section{Visualizing Pathological Descriptor}

According to Koch's Postulates, although the pathogen is invisible (at least to the naked eye), its properties could be determined by observing the symptoms on the subject after injecting the purified pathogen. Similarly, we evaluate and visualize the interpretative medical meaning of this descriptor by our novel Patho-GAN to generate fully controllable DR fundus images. Our goal is to synthesize the diabetic retinopathy fundus images (Fig~\ref{fig:koch}.(d)) that carry the lesions that appear on the pathological reference one (Fig~\ref{fig:koch}.(a)). Since our descriptor is lesion based, we could even arbitrarily manipulate the number and position of different symptoms. We show \niu{the overview of our pipeline in Fig~\ref{fig:pipeline_overview}, and details in Fig~\ref{fig:pipeline}}.

\subsection{Patho-GAN Overview}

As shown in Fig~\ref{fig:pipeline_overview}, the generator network synthesizes a retinal image \niu{with vessels in an segmentation map and controlled symptoms in descriptors}. To train the generator network, we add three sub-nets: the discriminator net, the perceptual net, and the DR detection net, as well as three related losses. The discriminator net \niu{is trained with the generator, distinguishing the synthesized images from the real ones to compete with and thus improves the generator}. To further enhance the pathological and  physiological details, we use the \niu{pre-trained} perceptual net to constrain the detail reconstruction. We also introduce \niu{pre-trained} DR detector to ensure the synthesized image exhibits the plausible symptoms. \niu{Different from the generator and discriminator, our perceptual net and DR detector are fixed and not updated during the training}. After the model is trained, the generator could easily obtain synthesized fundus in one forward pass given the vessel segmentation, descriptors, \niu{and a noise vector}.

\subsection{Generator Network}
Generator network $ G_\theta $ takes a vessel segmentation image $y \in \{ 0,1 \}^{W \times H}$, descriptors $ \mathcal{D}(x) $ and a noise code $z \in \mathbb{R}^Z $ as input, \niu{where $ y $ and $ \mathcal{D}(x) $ provide physiological and pathological information respectively, and $ z $ is used to provide randomness that models conditional image distribution. } The  output is  a synthesized diabetic retinopathy fundus image $\hat{x} \in [0,1]^{W \times H \times 3}$. 
The entire image synthesis process can be expressed as a function $\hat{x} = G_{\theta} ( y,\mathcal{D}(x),z )  $. 

As shown in Fig~\ref{fig:pipeline}, we use a U-shaped encoder-decoder network~\cite{ronneberger2015u} structure for the generator. The input vessel segmentation $ y $ is down-scaled by 6 blocks of Convolution-BatchNorm-LeakyRelu, with kernel size 4 \niu{(or 3)}, stride 2, and no pooling layers. The input descriptors $ \mathcal{D}(x) $ are used to reconstruct activation projections, which are then concatenated into the down-sampling process. The random noise $ z $ is fully connected and convolved into a block, and then concatenated into the bottleneck layer of the generator. For the up-scaling part, we choose to pile 6 blocks of Resize-Convolution-BatchNorm, rather than Transposed Convolution-BatchNorm in~\cite{zhao2018synthesizing} to avoid checkerboard artifacts~\cite{odena2016deconvolution}. To help retain details in the input segmentation and descriptors, skip connections from down-scaling features to up-scaling blocks are also considered in the network.

\subsection{Discriminator Network}

The discriminator net $ D_{\gamma} $ tries to distinguish the synthesized images from the real ones. Similarly, we can also define the discriminator as a discriminant function $p = D_{\gamma}( X , y)$, $p \in [0,1] $. When $X$ is the real image $x$, $p$ should tend to 1 and when $X$ is the composite image $\hat{x}$, $p$ should tend to 0. Structurally in Fig~\ref{fig:pipeline}, the detector is also built with 5 blocks of Convolution-BatchNorm-LeakyRelu, with kernel size 4 \niu{(or 3)}, stride 2, and no pooling layers.

\subsection{Loss Settings}

We follow the GAN's strategy and solve the  following optimization problem that characterizes the interplay between $ G_{\theta} $ and $ D_\gamma $:
\begin{align}
\label{align_allloss} \textstyle
\min_{\theta}  \textstyle\max_{\gamma} &L(G_{\theta}, D_{\gamma}) = \nonumber \\
\mathbb{E}_{x,y} [
	&L_\mathrm{adv}(x, \hat{x}, y; \theta, \gamma)
	 + w_p \cdot L_\mathrm{percept}(x, \hat{x}; \theta) \nonumber \\
	& + w_s \cdot L_\mathrm{severity} (x, \hat{x}; \theta)],
\end{align}
where $
L_\mathrm{adv} $ is the adversarial loss, with $ L_\mathrm{percept} $ and $  L_\mathrm{severity} $ being perceptual loss and severity loss.

\subsubsection{Adversarial Loss}
Adversarial loss is computed on real and synthesized images: $
L_\mathrm{adv} = \log D_\gamma(x, y) + 
\log(1 - D_\gamma(G_\theta(y,z),y))
$.
To be more specific, learning the discriminator parameter $ \gamma $ amounts to maximizing the adversarial loss $ L_\mathrm{adv} $.
As for the generator, parameter $ \theta $ is learned by minimizing a loss $ L_G = \tilde L_\mathrm{adv} + w_p \cdot L_\mathrm{percept} + w_s \cdot L_\mathrm{severity} $ where $ \tilde L_\mathrm{adv} $ is a simplified adversarial loss \niu{\sout{that are}} computed only on generated image:
\begin{align}
\label{align_Gloss} \textstyle
\tilde L_\mathrm{adv} = -\log D_\gamma (G_\theta(y, z),y) .
\end{align}

\subsubsection{Perceptual Loss}

% Though a  L1 loss (or MAE) between synthetic image could deliver a satisfactory result for style transfer application on common images, it fails to preserve the physiological details in the fundus. We will elaborate this in the experiment section. Therefore, we define \emph{retina detail loss} as:
Perceptual Loss is the divergence of real image and synthesized image measured in the feature space of the perceptual net. This is designed to help the  reconstruction of  both pathological and physiological details. 

We choose a pre-trained VGG-19~\cite{simonyan2015very} as the implementation of perceptual net. For specific layer $ \lambda $ and VGG feature extraction function $ F_V^\lambda $, we define \emph{perceptual loss} as:

\begin{align}
\label{align_percept_loss} 
L_\mathrm{percept} =  {\lVert F_V^\lambda(x) - F_V^\lambda(\hat x) \lVert}.
\end{align}

\subsubsection{Severity Loss}

The severity loss is introduced to make  the synthesized image $ \hat x $ medically equivalent to the reference real image $x$. Specifically, we constrain the synthesized image $ \hat x $ to have the same severity level with real image $ x $. The severity divergence is measured with trained DR detector o\_O, and the severity loss is defined as:
\begin{align}
	\label{align_severity_loss} 
	L_\mathrm{severity} = \|\text{DR}(x)-\text{DR}(\hat x)\|
\end{align}

\subsection{Implementation Details}

The chosen norm in above equations is L1. \niu{Noise dimension $Z$ is 400.} Weights for different losses are $ w_\mathrm{percept} = 1, w_\mathrm{severity} = 10 $. Based on experience, we set $ \lambda $ to be the second convolutional layer in the fourth block of VGG-19.

The batch size is set to 1. Before each training step, the input image values are scaled to $[-1, 1]$, and a random rotation is performed on the input. 
The training is done using the ADAM optimizer~\cite{kingma2015adam} and the learning rate is set to 0.0002 for the generator and 0.0001 for the discriminator. In order to ensure that generator and discriminator are adapted, we update generator twice then update discriminator once. During training, the noise code is sampled element-wise from zero-mean Gaussian with standard deviation 0.001; At testing run, it is sampled in the same manner but with a different standard deviation of 0.1.
The training finishes after 1000 epochs.

\hfill

\section{Experiment Results}
   	
\subsection{Datasets}
There are several datasets of fundus image online available, such as DRIVE~\cite{staal:2004-855},  Kaggle (EyePACS)~\cite{kaggle2015diabetic}, IDRiD~\cite{porwal2020idrid}, Retinal-Lesions~\cite{wei2019retinal-lesions} and FGADR~\cite{zhou2020benchmark}. 
DRIVE~\cite{staal:2004-855} contains 40 fundamental retinal images with pixel-level vessel segmentation, but there are few lesions on images. The Kaggle (EyePACS)~\cite{kaggle2015diabetic} contains 88.7k retinal images of 5 levels DR grades (including non DR). Each image is labelled with a DR grade. To further boost the DR research, IDRiD~\cite{porwal2020idrid} provides pixel-wise segmentation of 4 types of lesions on 81 images. Retinal-Lesions datasets~\cite{wei2019retinal-lesions} annotates up to 8 kinds of lesions with circle regions. Recently, FGADR~\cite{zhou2020benchmark} released 1842 images with pixel-wise segmentation of 6 kinds of lesions.
We summarize the attributes of datasets in TABLE~\ref{tab:datasets}.% (In Retiinal-Lesions, and FGADR, we only choose some of the images during primary evaluation. We'll report full result in the revised version.) %\textcolor{red}{please explain why not all images were used to train and test (Retiinal-Lesions, and FGADR)}

\begin{table*}[h!]
	\caption{Retinal image datasets and their attributes.}
	\label{tab:datasets}
	\begin{center}
		\begin{tabular}{ccccc}
			\toprule
			Dataset Name & \multicolumn{1}{c}{DR Lesion Annotation} & Vessel Segmentation & Resolution & Number of Images \\
			\midrule
			DRIVE~\cite{staal:2004-855} & \multicolumn{1}{c}{(No DR)} & Yes   & 584$\times$565 & 20 train + 20 test \\
			EyePACS Kaggle~\cite{kaggle2015diabetic} & \multicolumn{1}{c}{Only severity levels} & No    & 1444$\times$1444$\sim$2184$\times$3456 & 35.1k train + 53.6k test \\
			IDRiD~\cite{porwal2020idrid} & \multicolumn{1}{c}{Pixel-wise lesion segmentation} & No    & 4288$\times$2848 & 54 train + 27 test \\
			Retinal-Lesions~\cite{wei2019retinal-lesions} & Lesion annotation in circle \& severity levels & No    & 896$\times$896 & \niu{337 train + 1256 test} \\
			FGADR~\cite{zhou2020benchmark} & Pixel-wise lesion segmentation \& severity levels & No    & 1280$\times$1280 & \niu{500 train + 1342 test} \\
			\bottomrule
		\end{tabular}%
	\end{center}

\end{table*}

In our experiments, DRIVE images and the vessel segmentation is used for training a vessel segmentation network, SA-UNet\cite{guo2020saunet}. Kaggle images and severity labels are used to train the o\_O DR detector. We then train and evaluate our  Patho-GAN on  IDRiD~\cite{porwal2020idrid},  Retinal-Lesions ~\cite{wei2019retinal-lesions} and  FGADR~\cite{zhou2020benchmark}.

\subsection{Synthesizing Images from Descriptors}

\niu{Since the  images of IDRiD, Retinal-Lesions, FGADR contain a great variety of DR lesions, we use their retinal images for training Patho-GAN. The images are partitioned into train/test sets as described in TABLE~\ref{tab:datasets}. Before our experiments, the dataset images  are firstly downscaled to 512 $\times$ 512 real images $x$. For the interpretation of the pre-trained DR detector, descriptors $ \mathcal{D}(x) $ are extracted from corresponding retinal image by o\_O DR detector, without accessing any lesion annotations in datasets. Since vessel annotation is not provided by these datasets, we compute the related binary vessel segmentation $y(x)$ through SA-UNet\cite{guo2020saunet} pre-trained on lesion-free DRIVE dataset. The 400-dimension noise $z$ is randomly sampled where $z_i \sim N(0, 0.001)$. With real images $x$ in the train-set, vessel segmentation $y(x)$ and noise $z$, we trained 3 Patho-GAN models on IDRiD, Retinal-Lesions, and FGADR respectively. 
}

\begin{figure}[t]
	\begin{center}
		\includegraphics[width=\columnwidth]{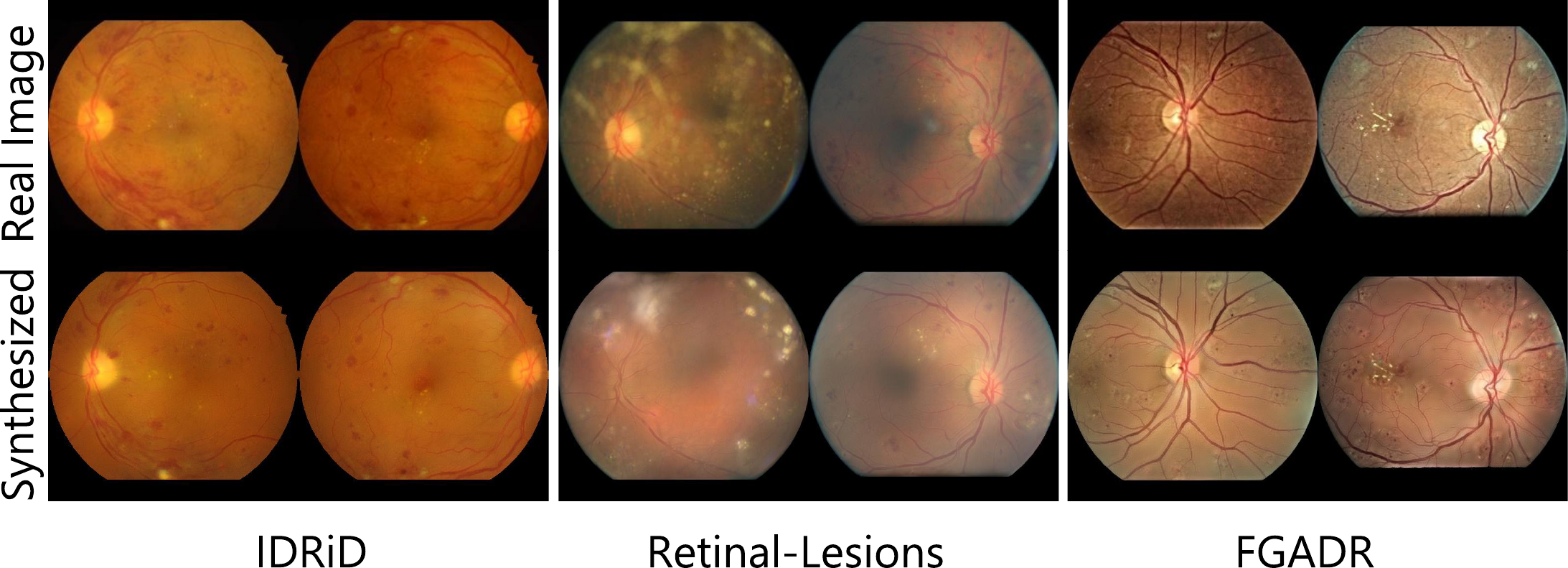}
	\end{center}
	\caption{\niu{Real images (row 1) and synthesized ones by our Patho-GAN (row 2) on three datasets. Boundary mask is applied on synthesized results.}}
	\label{fig:exp_synthesized_results}
\end{figure}

\niu{In testing phase, with test-set image $ x $ as a reference, we input $ \mathcal{D}(x) $, $y(x)$ and $z$ into trained generators to get synthesized fundus $ \hat{x} $. Same with \cite{zhao2018synthesizing,Niu19AAAI,Zhou19DRGAN}, in order to keep same field of view (FOV), the generated images are cropped with the same FOV boundary of the reference image $ x $ before visualization. Some pairs of real image $ x $ and generated image $ \hat{x} $ are presented in Fig~\ref{fig:exp_synthesized_results}, showing that} Patho-GAN can reconstruct a variety of photo-realistic lesions, including microaneurysms, hemorrhages, soft and hard exudates, with the similar locations and appearance.

\subsection{Lesion Manipulations}
\label{sec:lesion_manipulations}

Since our descriptors are designed as lesion-based with spatial coordinates, we could manipulate lesions' positions by modifying their coordinates, and manipulate lesions' numbers by cloning or removing specific descriptors. Here, we demonstrate some results.

%To further demonstrate that the Patho-GAN generated pathological images based on these descriptors that encode the lesions, we try to edit the input descriptors and show how the generated images change.

%and see if the synthesized lesions are also changed.

    \subsubsection{Lesion Relocation} We could arbitrarily  relocate individual lesion to any position by controlling the coordinates (left and top distance) of descriptors. The \niu{modified descriptor layout and synthesized images} are shown in row 1 and 2 of Fig~\ref{fig:exp_manipulation}. Note that images in each group are generated with same vessel segmentation and descriptors as input, except that the descriptor coordinates are different. We could see the lesions relocate their positions according to  the given coordinates.
    %As the figure indicates, the position of generated lesions get also relocated, as is controlled by the input descriptors.

    \subsubsection{Number Manipulation} 
\begin{figure}[t]
	\begin{center}
		\includegraphics[width=\columnwidth]{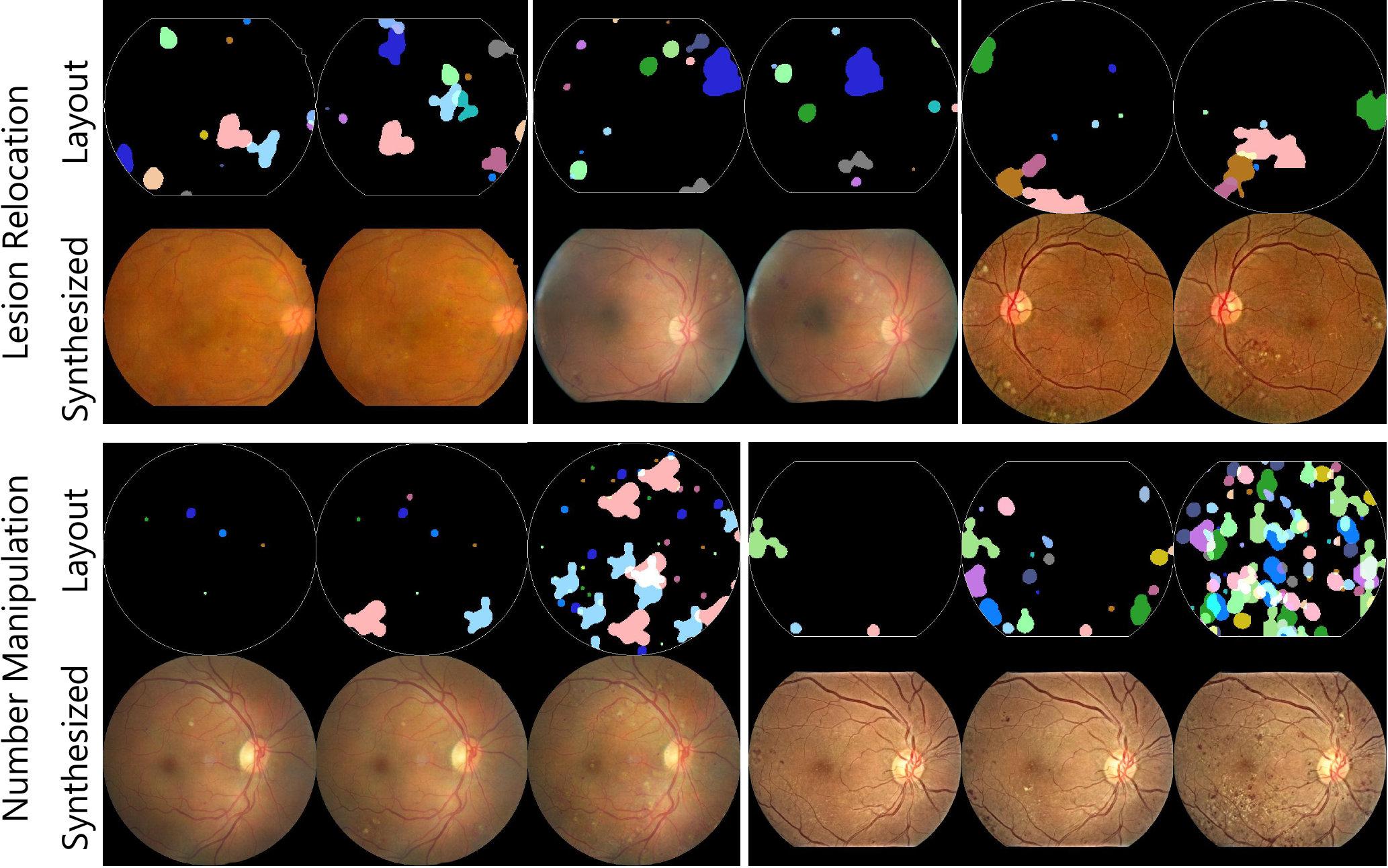}
	\end{center}
	\caption{\niu{Results of lesion manipulation. The layout map shows the distribution of descriptors, whose color means the descriptor's identity. By modifying descriptors, we can manipulate the synthesized lesions.} }
	\label{fig:exp_manipulation}
\end{figure}

By cloning or removing some of the input descriptors, we can control the amount of synthesized lesions, \textit{i.e.}, decreasing or increasing their number. Row 3 and 4 of Fig~\ref{fig:exp_manipulation} demonstrate the result of lesion number manipulation. \niu{In each group shown here, we generate lesions of 3 different amount: less, middle, and more lesions. Middle lesions are synthesized with original descriptor layout. We randomly remove descriptors in the original layout to generate less lesions. We clone each of the descriptors by several times and randomly distribute them to generate more lesions.} 
    
\begin{figure}[t]
	\begin{center}
		\includegraphics[width=\columnwidth]{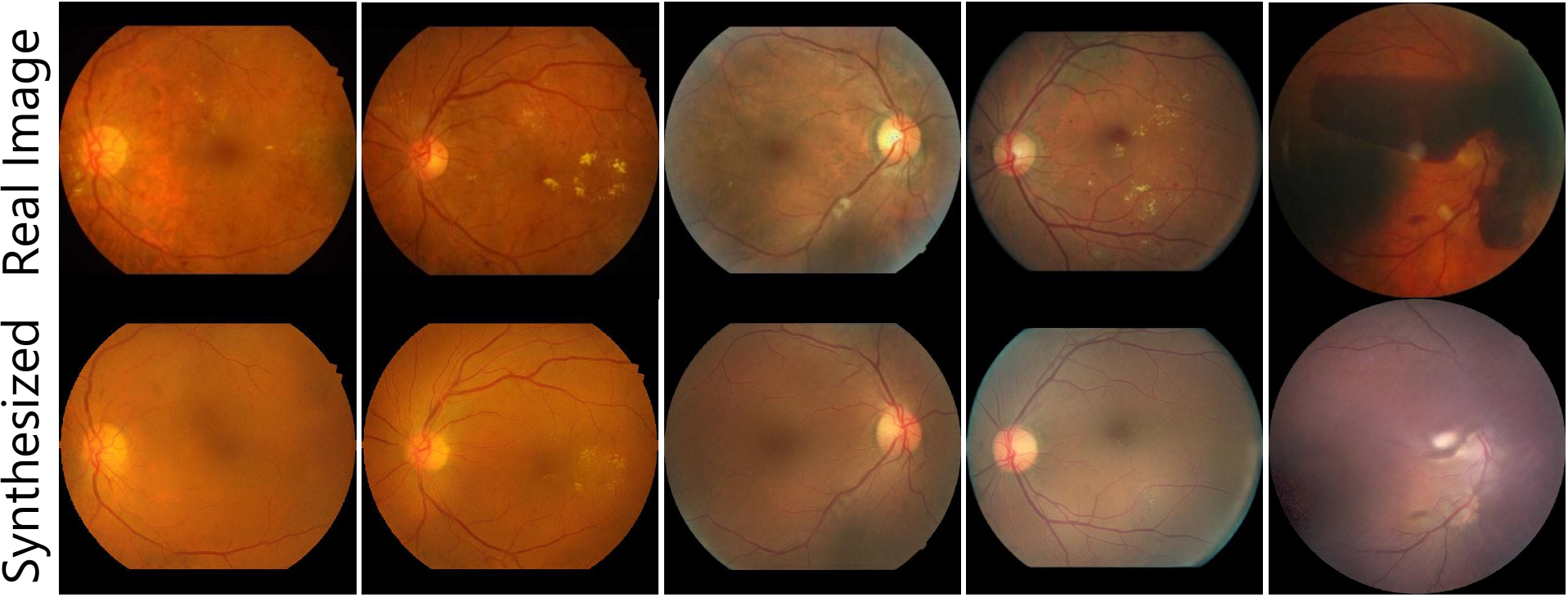}
	\end{center}
	\caption{\niu{With empty pathological descriptors, Patho-GAN actually eliminates lesions in the reference image}}
	\label{fig:exp_eliminating}
\end{figure}

    An interesting observation is that we could even eliminate existing lesions in a DR retinal image by inputting no (empty) pathological descriptors. As shown in Fig~\ref{fig:exp_eliminating}. Row 1 are the original retinal images with DR, while row 2 are synthesized with no descriptors. The pathological symptoms in row 1 almost disappeared in row 2.
%    \subsubsection{Category Manipulation} We collected some descriptors extracted from the same type of lesions, such as microaneurysms, hemorrhages, soft exudates and hard exudates, and feed them into Patho-GAN with same vessel segmentation. As show in row 5 and 6 of Fig~\ref{fig:exp_manipulation}, the output image contains only the same type of lesions.

%Above experiments have shown that our method is able to synthesize the image with the controllable symptoms in terms of positions, numbers and appearances.

%positions, numbers and appearances of symptoms that are directly

%and our synthesized images carry the symptoms that are directly related to diabetic retinopathy diagnosis.

\subsection{Medical Interpretation}

We also evaluate the synthesized images on the o\_O DR detector~\cite{oO2016detector} to evaluate its DR severity on the image level.  According to~\cite{aao2002drscale}, number of lesions like microaneurysms is an important criteria for the severity diagnose. Here we manipulate the number of lesions on retinal images and report the severity prediction by o\_O DR detector~\cite{oO2016detector}.

\begin{figure}[t]
	\begin{center}
		\includegraphics[width=\columnwidth]{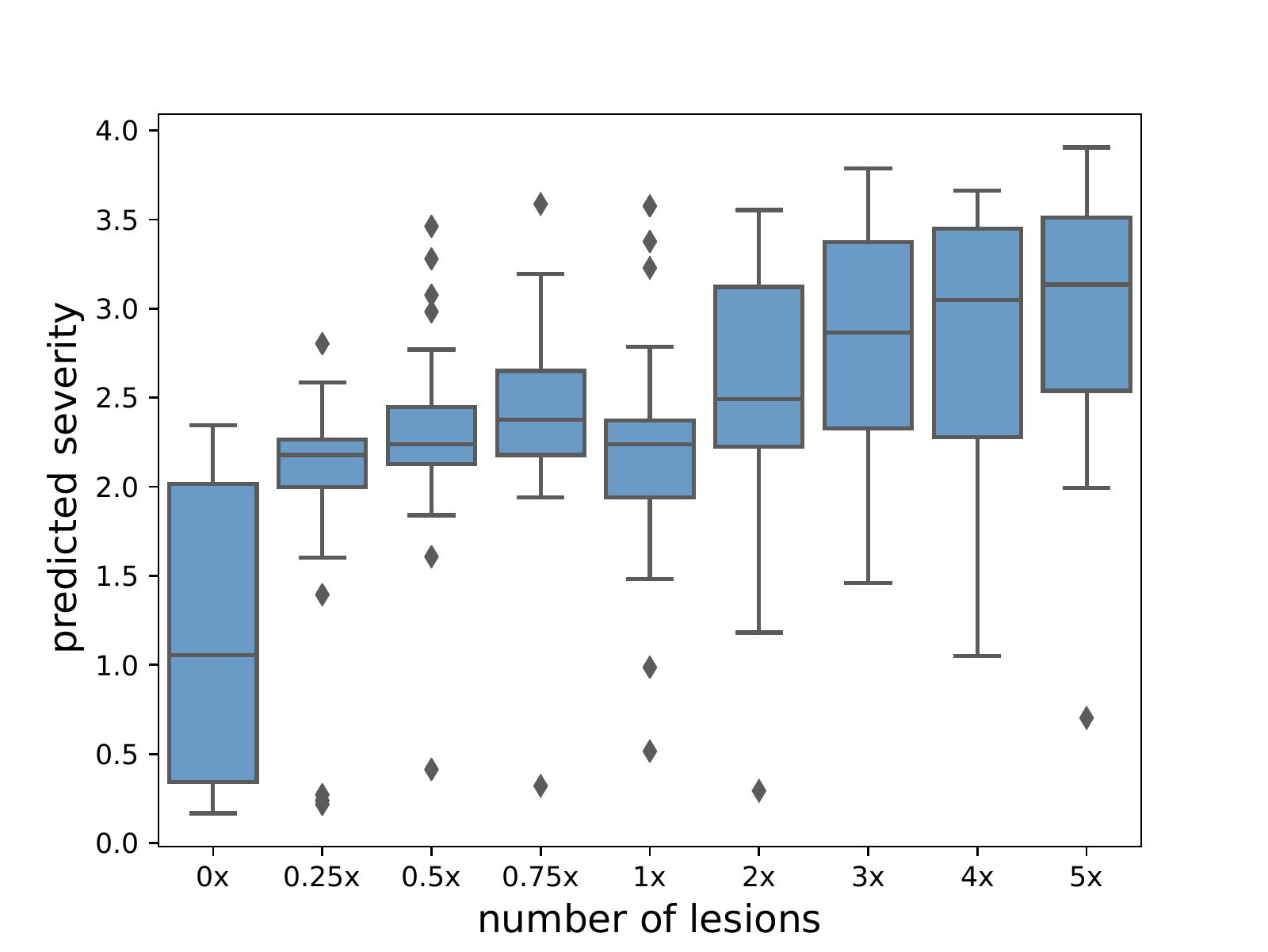}
	\end{center}
	\caption{The severity score goes up with increasing lesions number.}
	\label{fig:number_severity}
\end{figure}

\niu{IDRiD consists of 27 test images with 4 type of lesions including microaneurysms, hemorrhages, hard and soft exudates, and each image has enough and even amount of DR lesions. So we choose IDRiD for this experiment. Using the number manipulation method described in last section, we modify each IDRiD test image $ x $ by removing or multiply the original descriptors $ \mathcal{D}(x) $, generating different amount of lesions varying from 0 to 5 times of the original number. Specifically, we randomly remove 100\%, 75\%, 50\%, 25\%, 0\% from $ \mathcal{D}(x) $ to generate image with less (0x, 0.25x, 0.5x, 0.75x) or equal (1x) lesions. We multiple descriptors in this image $ \mathcal{D}(x) $ by 2 to 5 times to generate more (2x, 3x, 4x, 5x) lesions. The statistics of predicted DR severity are reported in the box plot Fig~\ref{fig:number_severity}. As a result, by increasing the lesion number, o\_O DR detector~\cite{oO2016detector} also returns a result with increasing severity, consistent with the medical interpretation~\cite{aao2002drscale}.
}

%the medical knowledge: the severity score goes up with increasing lesions number.

\subsection{Qualitative Evaluation}

\begin{figure}[t]
	\begin{center}
		\includegraphics[width=\columnwidth]{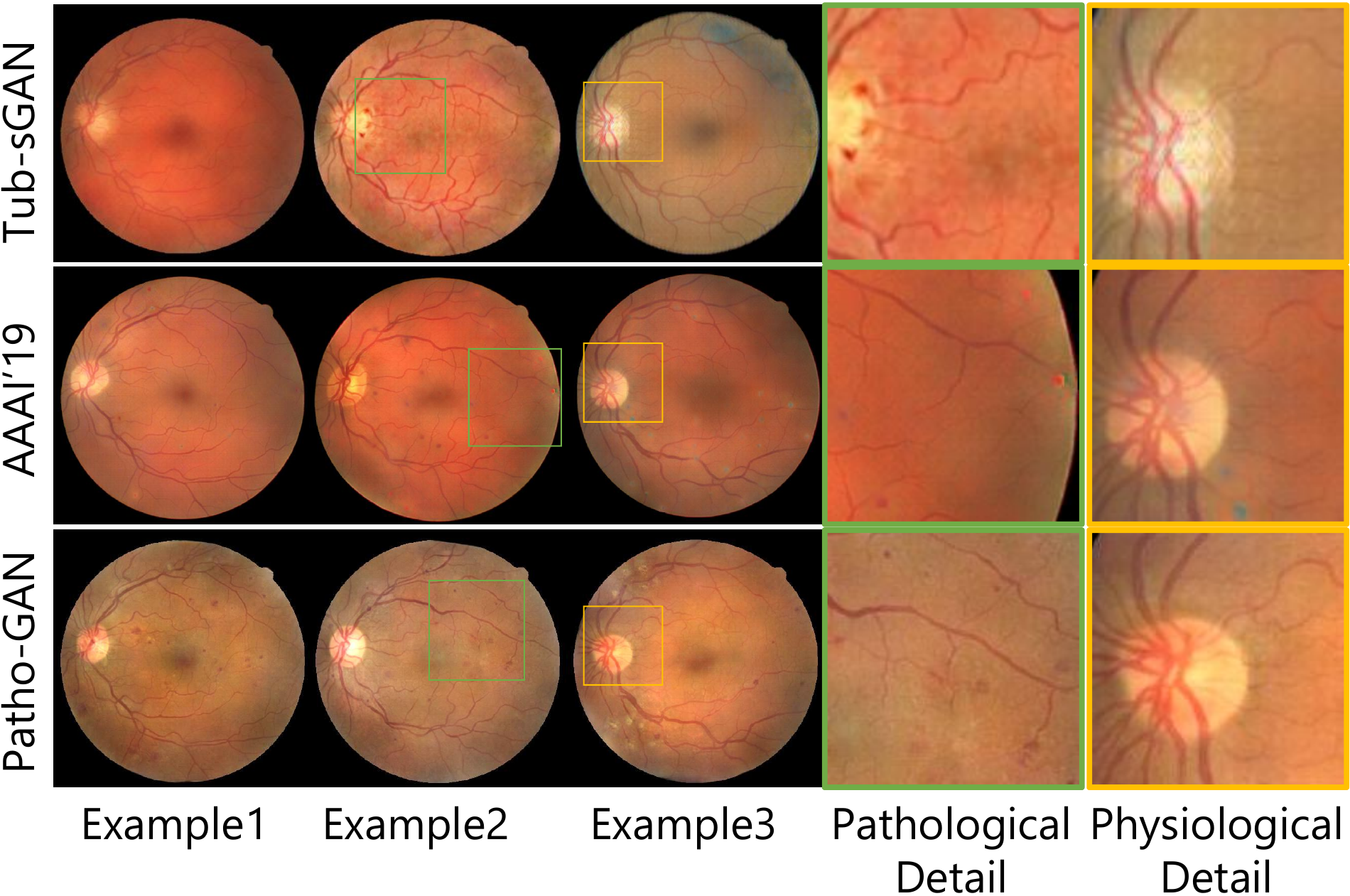}
	\end{center}
	\caption{\niu{Comparison of different DR image generation methods. Our synthesized images (row 3) are more photo-realistic in both pathological symptoms and physiological details.}}
	\label{fig:exp_qualitative}
\end{figure}

In this subsection, we compare our Patho-GAN together with two related generation methods,  Tub-sGAN~\cite{zhao2018synthesizing} and our previous work ~\cite{Niu19AAAI} in Fig~\ref{fig:exp_qualitative}. \niu{Pathologically (in col 4), Tub-sGAN can not synthesize lesions, or gets wrong lesions, while  ~\cite{Niu19AAAI}  generates scattered and sometimes unreal lesions. However, our Patho-GAN  synthesizes realistic lesions with various visual appearances. Physiologically (in col 5), our generated results have clear boundaries of optic disc, and get rid of checkerboard artifacts (repeated grid patterns) mainly caused by transposed-convolution up-sampling. } As a conclusion, the images we generate are  more photo-realistic in both pathological symptoms and physiological details.

\subsection{Quantitative Evaluation}

Here we also perform quantitative evaluation on Patho-GAN and other related methods~\cite{zhao2018synthesizing}~\cite{Niu19AAAI}. We generate images using these methods on IDRiD, Retinal-Lesions and FGADR, and evaluate the similarity between generated images and real images using Freshet Inception Distance (FID) and Mean Squared Error (MSE).
Freshet Inception Distance (FID) is usually adopted to statistically measure the divergence of two image sets, while Mean Squared Error (MSE) measures the pixel-wise average difference of two images or one-to-one related image sets.
The results in TABLE~\ref{tab:quantitative_comparison} \niu{indicate that we have better score than previous works, and that Patho-GAN generates more similar retinal images to real images. }

\begin{table}[t]
	\caption{Quantitative measurement on different methods (lower is better).}
	\label{tab:quantitative_comparison}	
	\begin{center}
\makebox[\columnwidth][c]{%
% Table generated by Excel2LaTeX from sheet 'FID (6.9)'
%\begin{tabular}{c|cc|cc|cc}
\begin{tabular}{>{\color{revcolor}}c|>{\color{revcolor}}c>{\color{revcolor}}c|>{\color{revcolor}}c>{\color{revcolor}}c|>{\color{revcolor}}c>{\color{revcolor}}c}
	\toprule
	\multirow{2}[2]{*}{Methods} & 
	\multicolumn{2}{c|}{\color{revcolor}IDRiD} & 
	\multicolumn{2}{c|}{\color{revcolor}Retinal-Lesions} & 
	\multicolumn{2}{c}{\color{revcolor}FGADR} \\
	%\multicolumn{2}{c|}{IDRiD} & \multicolumn{2}{c|}{Retinal-Lesions} & \multicolumn{2}{c}{FGADR} \\
	& FID$\downarrow$  & MSE$\downarrow$  & FID$\downarrow$  & MSE$\downarrow$  & FID$\downarrow$  & MSE$\downarrow$ \\
	\midrule
	Tub-sGAN~\cite{zhao2018synthesizing} & 111.36  & 0.0119  & 69.42  & 0.0250  & 40.67  & 0.0154  \\
	~\cite{Niu19AAAI} & 117.82  & 0.0128  & 40.76  & 0.0214  & 95.06  & 0.0146  \\
	\midrule
	Patho-GAN$_\mathrm{4\times4}$  & 81.16  & 0.0093  & \textbf{22.28} & \textbf{0.0144} & \textbf{20.34} & 0.0115  \\
	Patho-GAN$_\mathrm{3x3}$ & \textbf{80.13} & \textbf{0.0086} & 24.37  & 0.0149  & 21.11  & \textbf{0.0107} \\
	\bottomrule
\end{tabular}%
}

{ \vspace{0.5em} * \niu{Subscription is the convolution kernel size ($4\times4$ or $3\times3$) of Patho-GAN.}}
	\end{center}

\end{table}

\subsection{Ablation Study of Perceptual Loss}

To evaluate the  perceptual loss on the reconstruction of both pathological and physiological details, we conduct the ablation analysis of perceptual loss by adjusting its weights $ w_p $. 

\begin{figure}[t]
	\begin{center}
		\includegraphics[width=0.95\columnwidth]{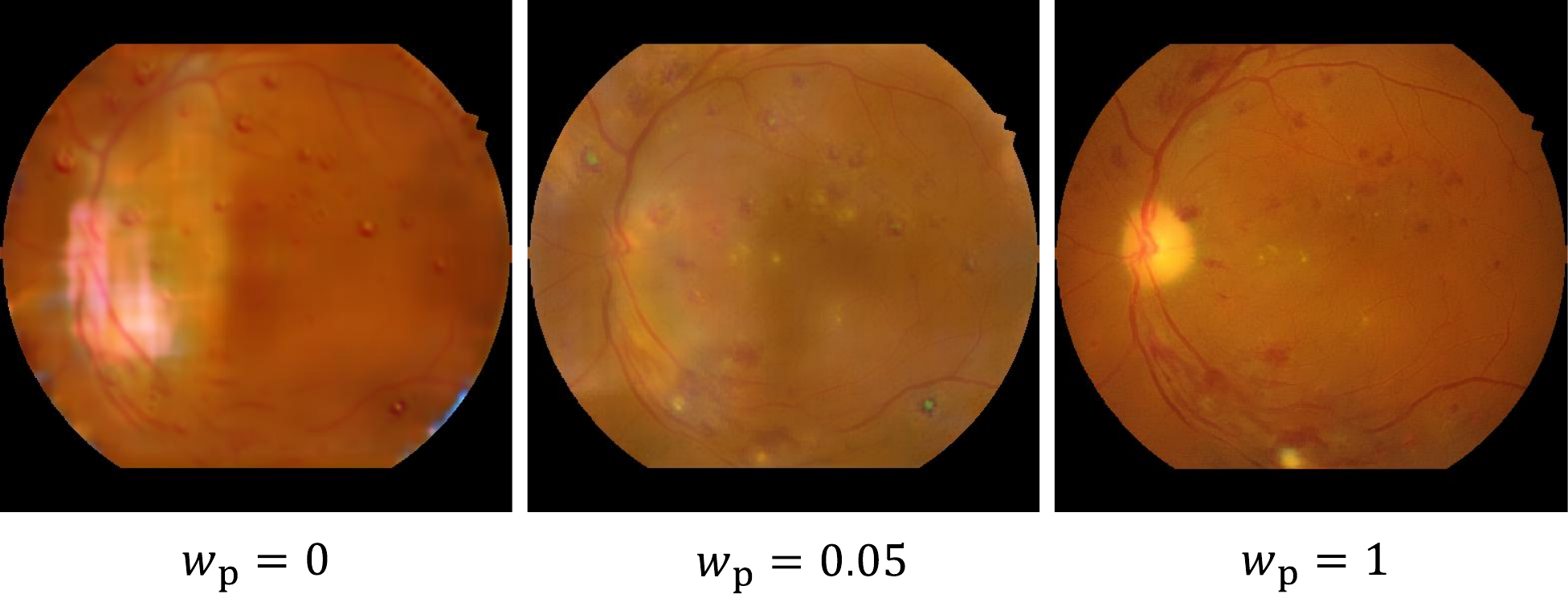}
	\end{center}
	\caption{The generated image with increasing weights of perceptual loss.}
	\label{fig:ablation_percept}
\end{figure}

As shown in Fig~\ref{fig:ablation_percept}, when reducing the weight of perceptual loss, the synthesized vessels and optic discs start to blur. The generated lesions become unreal while the background noises in the image increase. In summary, perceptual loss plays a key role in restraining the generator to learn pathological and physiological details effectively.

\subsection{Computational Efficiency}

To synthesize images of a new style or descriptors, both Tub-sGAN~\cite{zhao2018synthesizing} and previous method~\cite{Niu19AAAI} need to train a whole new generator network with the new lesion style or layout  in the training loss. This usually takes 0.5 - 2 hours using 1 GPU of Nvidia Titan Xp.

%because the lesion style or layout is used in loss function and works in training phase.

By taking descriptors as network input, Patho-GAN can generate images with new descriptors through the inference in less than 1 second. The speedup endows our method the potential for data augmentation in medical imaging analysis.

\section{Discussion}

\subsection{Interpretability obtained with Patho-GAN}
\niu{Here we briefly talk about the interpretability attained with our framework.}

\niu{{The DR detector focus on lesions.}} In this paper, we try to interpret deep learning DR detector using a methodology like Koch's Postulates illustrated in Fig \ref{fig:koch}. We first extract pathological descriptors from a retinal image to describe the positions, dimensions and appearances of DR lesions, and then prove the effectiveness of descriptors by reconstructing lesions \niu{from them faithfully. According to the Koch's Postulates we conclude that the neuron activate pattern is closely related with the pathological lesions.}

\begin{figure}[t]
	\centering
	\includegraphics[width=\columnwidth]{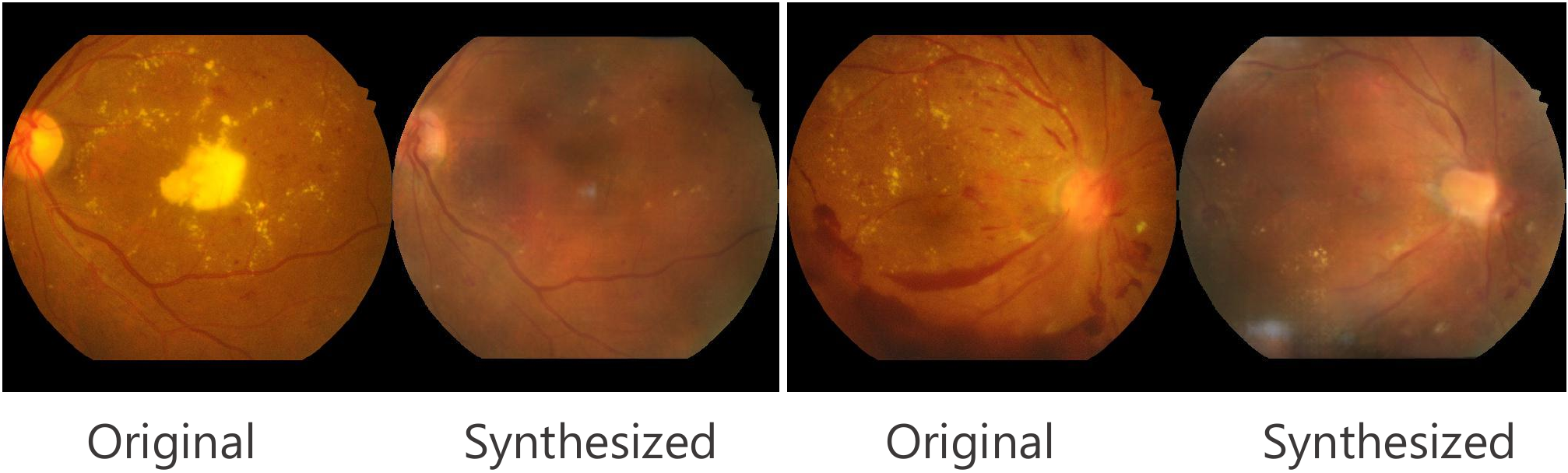}
	\caption{When there are large-region lesions in the reference images, Patho-GAN left out large ones and only synthesize small ones. In other words, Patho-GAN provide a perspective of how the DR detector perceive an retinal image. }
	\label{fig:large_lesion}
\end{figure}

The visualized activation projections in Fig~ \ref{fig:activation_projection} and Fig~\ref{fig:detected_lesion} show high correlation with the original images in lesion areas, and moreover it is able to locate lesions in an unsupervised manner. 
%We don't use manually labeled pixel-wise lesion masks like \cite{Zhou19DRGAN} but detect lesions automatically with the DR detector and corresponding activation network to make sure that the extracted descriptors keep a consistency on detect-ability with the DR detector. 
\niu{As Fig~\ref{fig:activation_projection_failure} indicates, large-region lesions are not distinguishable from the activation projections as DR detector is able to make the diagnose only based on microaneurysms. Patho-GAN faithfully interprets the network by only generating small pathological symptoms (microaneurysms) without synthesizing  large pathological symptoms (exudates or hemorrhage) in Fig~\ref{fig:large_lesion}. Accordingly, Patho-GAN provides a perspective on how the o\_O DR detector perceive an retinal image: it can predict severity without detecting large lesions.}

%the DR detector doesn't recognize all lesions. Therefore, the reconstruction from descriptors explains the detect-ability of the DR detector.

\niu{{The DR detector is sensitive to various kinds of lesions.}} As shown in Fig \ref{fig:exp_synthesized_results}, the retinal image reconstruction method Patho-GAN could generate different types of lesions. And further proven by lesion manipulations in Fig~\ref{fig:exp_manipulation}, the generated lesions are synthesized from pathological descriptors. \niu{To conclude, descriptors contain information of various kinds of lesions, leading to the explanation that the DR detector is sensitive to various lesions.}

\niu{{The DR detector is consistent with the medical interpretation.}} In the experiment of medical interpretation, we tried to generate retinal images of different lesion counts and feed them into DR detector. \niu{The results in Fig~\ref{fig:number_severity} demonstrate that, when introducing more lesions, the o\_O DR detector~\cite{oO2016detector} returns a result with increasing severity consistent with the medical interpretation~\cite{aao2002drscale}.
}

\niu{
Apart from the interpretability, Patho-GAN itself is superior to existing generating methods on the qualitative evaluation, quantitative evaluation, and computational efficiency. Fig~\ref{fig:exp_qualitative} compares retinal images generated using Patho-GAN and related methods. Patho-GAN generates high-quality images with realistic lesions and without checkerboard artifacts. Table~\ref{tab:quantitative_comparison} shows that Patho-GAN generates more similar images to the real ones in the sense of FID and MSE. Finally, the computational speed is also much faster compared to \cite{Niu19AAAI}.}

\subsection{Network Architecture and Parameter Settings}

\niu{The medical interpretation result in Fig~\ref{fig:number_severity} proves the effectiveness of severity loss in keeping grades. The ablation study in Fig~\ref{fig:ablation_percept} implies the importance of perceptual loss in preserving physiological and pathological structures. }

%But not all structures or settings can be works as well, or vital to the performance.
 
\niu{
Gaussian noise code z is used in CGAN~\cite{mirza2014conditional}  to model conditional distributions of the real data that avoids deterministic output. When it comes to image synthesis, as pix2pix~\cite{isola2017image} finds, generator simply learns to ignore the noise. In our experiments, noise z mainly contributes to the brightness of retinal background, and contributes not much to the stochasticity and variety of physiological structures and lesions. The perfect reconstruction of retinal background colors and appearance needs further research.
}

\niu{
Most of the parameter settings are inhered from our previous work~\cite{Niu19AAAI}, or chosen empirically. For example, the convolution kernel size of $ G $ and $ D $. The setting $ 4\times4 $ is inherited from Tub-sGAN~\cite{zhao2018synthesizing} and originated from pix2pix~\cite{isola2017image} and DCGAN~\cite{radford2016unsupervised}, which is only a convention in GAN and style transfer area. We have tried to train Patho-GAN with kernel of $ 4\times4 $ and $ 3\times3 $. The FID and MSE scores between real images and generated images are reported in TABLE~\ref{tab:quantitative_comparison}. Patho-GAN trained with different kernel sizes get similar better scores compared to previous method Tub-sGAN~\cite{zhao2018synthesizing} and AAAI'19~\cite{Niu19AAAI}. Different kernel sizes work equally well.
}

\subsection{Future Works}
%\niu{
%Patho-GAN only uses retinal images in IDRiD, Retinal-Lesions, and FGADR. As shown in TABLE~\ref{tab:datasets}, these datasets also provide lesions annotations at pixel level or rough circle boundaries. With these information, the DR detector's interpretability ability may be measured quantitatively in future works. 
%}

%For example, we could show show the consistency of the descriptor layout and lesion segmentation. 

\niu{
In future works, our interpretation framework is not limited to medical tasks such as DR detection. It also applies on  broader critical applications such as medicine, banking and self-driving automobiles. Our method has  the potential to enhance decision reliability, algorithm fairness, and ensure network performance of general CNN models.
}

\section{Conclusion}
To exploit the network interpretability in medical imaging, following similar methodology to Koch's Postulates, we propose a novel strategy to encode pathological descriptors from the activated neurons directly related to the prediction, and a GAN based visualization method, Patho-GAN, to visualize the pathological descriptor into a pathology retinal image from an unseen binary vessel segmentation. \niu{The images we generated have shown medical plausible and controllable symptoms, proving that specific lesions are directly related to the DR grade prediction.} This explainable work helps medical community to further understand on how deep learning makes prediction and encourage more collaboration.

\section*{Acknowledgment}

This work was supported by National Natural Science Foundation of China (NSFC) under Grant 61972012 and JST, ACT-X Grant Number JPMJAX190D, Japan.

\bibliographystyle{IEEEtran}
\bibliography{mybib}

\end{document}